\magnification=1200
\hsize=16truecm
\vsize=24truecm
\baselineskip 18 truept
\voffset=-0.5truecm
\overfullrule=0pt

\def\Ai{\hbox{\hbox{${\cal A}$}}\kern-1.9mm{\hbox{${/}$}}}
\def\Vi{\hbox{\hbox{${\cal V}$}}\kern-1.9mm{\hbox{${/}$}}}
\def\Di{\hbox{\hbox{${\cal D}$}}\kern-1.9mm{\hbox{${/}$}}}
\def\lam{\hbox{\hbox{${\lambda}$}}\kern-1.6mm{\hbox{${/}$}}}
\def\D{\hbox{\hbox{${D}$}}\kern-1.9mm{\hbox{${/}$}}}
\def\A{\hbox{\hbox{${A}$}}\kern-1.8mm{\hbox{${/}$}}}
\def\V{\hbox{\hbox{${V}$}}\kern-1.9mm{\hbox{${/}$}}}
\def\parz{\hbox{\hbox{${\partial}$}}\kern-1.7mm{\hbox{${/}$}}}
\def\B{\hbox{\hbox{${B}$}}\kern-1.7mm{\hbox{${/}$}}}
\def\R{\hbox{\hbox{${R}$}}\kern-1.7mm{\hbox{${/}$}}}
\def\si{\hbox{\hbox{${\xi}$}}\kern-1.7mm{\hbox{${/}$}}}

\null
\hskip 10truecm Preprint Padova, DFPD 97/TH/10

\hskip 10truecm hep-th/9703xxx

\hskip 10truecm March, 1997

\vskip 1truecm

\centerline{\bf On the Equivalence of Different Formulations}

\centerline{\bf of the M Theory Five--Brane}

\vskip 2.5truecm
\centerline{Igor Bandos$^1$, Kurt Lechner$^2$, Alexei Nurmagambetov$^1$,}

\centerline{Paolo Pasti$^2$, Dmitri Sorokin$^1$ and Mario Tonin$^2$}

\vskip 1.5truecm
\centerline{\it $^1$ National Science Center, Kharkov Institute of Physics
and Technology,}

\centerline{\it Kharkov, 310108, Ukraine} \vskip 0.3truecm

\centerline{\it
$^2$ Universit\`a degli Studi di Padova, Dipartimento di Fisica ``G.Galilei"}

\centerline{\it ed INFN, Sezione di Padova, Via F.Marzolo, 8, 35131 Padova,
Italy}

\vskip 2truecm

\centerline{\bf ABSTRACT}

We show  that the field equations for the supercoordinates and the
self--dual antisymmetric tensor field derived from the recently
constructed $\kappa$--invariant action for the M theory five-brane are
equivalent to the equations of motion obtained in the doubly
supersymmetric geometrical approach at the worldvolume component level.

\vskip 0.5truecm

\vfill\eject

A super--five--brane sigma model in eleven dimensions is an essential
ingredient of $M$--theory, an eleven--dimensional theory that has been
conjectured in the framework of string duality.
\par
This brane contains on its worldvolume a self--dual tensor whose presence
hampered for a long time a complete description of this extended object due
to known problems with (manifest) Lorentz invariance. Only
very recently consistent worldvolume formulations of
the super--five--brane in $D=11$ have been proposed.
\par
A complete covariant Dirac--Born--Infeld (DBI)--like
action for the bosonic $D=11$ five--brane was constructed in [1], and in
[2,3]
it was generalized to a $k$--invariant action for the M--theory
super--five--brane in $D=11$ supergravity backgrounds.
The construction is based on a previous knowledge about the structure
of different parts of the action obtained in [4,5].
A manifestly Lorentz invariant treatment of the self--dual field was achieved
by applying a method proposed in [6].  The method consists in
introducing an auxiliary scalar field which does not propagate and can be
eliminated by fixing a new local symmetry of the action at the expense of
losing manifest Lorentz invariance.
The five--brane action in this gauge was considered in detail in [3].
An advantage of having the covariant formulation is
that the five--brane action has a conventional worldvolume diffeomorphism
invariant form, which simplifies the analysis of its structure and relation
with other extended objects, for instance,  with a heterotic string [7].

The presence of the
auxiliary field also reveals nontrivial topological properties of the model.
\par
At first sight a completely different approach is that of Refs. [8,9].
It is based on a doubly supersymmetric geometrical approach to
describing super--p--brane dynamics [10--14] (and refs. therein),
where both
the five--brane worldvolume and the eleven--dimensional target space are
superspaces and what one gets are geometrical conditions
specifying a superembedding of worldvolume superspace into
a target superspace. In many cases, as it happens with the five--brane,
these conditions put the theory on the mass shell
and yield
superfield equations of motion without any knowledge on the
structure of the five--brane action. Thus, this method does not furnish
any action from which these equations can be derived.

\par
In this letter we shall prove that the
approaches of [1,2,3] and that of [8,9] are equivalent in the sense that the
worldvolume component equations of motion (for $X$, $\vartheta$ and the
chiral tensor) in the latter are equivalent to the equations of motion
derived from the action in the former.  This result is not so surprising,
since, after a double dimensional reduction [1,9,15], both formulations yield
a DBI--like structure of the D--four--brane in $D=10$ and, moreover, as
shown in [16], the field equations of the chiral tensor are the same in
the both
cases.  However, in Ref. [9] the authors presented $\kappa$--variations of
their component fields which look rather different from those in the action
approach. Here we will show that the two sets of $\kappa$--transformations
are in fact related by a redefinition of the parameter $\kappa$.
\par
The paper is organized as follows.
We start with a review of the structure of
the five--brane action, then consider the five--brane component equations
obtained by use of the geometrical approach, and finally derive a set of
relations and identities which establish the equivalence between the two
formulations.
\par
The five--brane is described by the supercoordinates
$$
Z^{\underline{M}} (x) \equiv \left(X^{\underline{m}},
\vartheta^{\underline{\mu}}\right) \ (\underline{m} = 0,..., 10; \
\underline{\mu} = 1,...,32).  $$ together with the worldvolume $2-$form
$A_2(x)$ which is the potential of a self--dual tensor. The $x^m$ are the
coordinates of the brane worldvolume (underlined indices refer to the
eleven--dimensional target superspace while indices which are not underlined
refer to the six--dimensional worldvolume). The curved target superspace
background is specified by the supervielbeins $E^{\underline{A}} (Z) \equiv
dZ^{\underline{M}} E_{\underline{M}}{}^{\underline{A}} (Z)$ and the Lorentz
superconnection $\Omega_{\underline{A}}{}^{\underline{B}}(Z)$ with torsion
$T^{\underline{A}} = D E^{\underline{A}}$ and Lorentz curvature
$R_{\underline {A}}{}^{\underline{B}} (Z)$, together with the target space
three--superform $C_3 (Z)$  and six--superform $C_6 (Z)$ with curvatures $$
\eqalignno { R_4& = dC_3&(1)\cr R_7& = dC_6 + {1\over 2} C_3 \wedge
R_4.&(2)\cr } $$ $C_3$ and $C_6$ are dual potentials in the sense that $$
R_{\underline {a}_1...\underline{a}_7}= {1 \over 4!}
\epsilon_{\underline{a}_1...\underline{a}_7}{}^{\underline{b}_1...
\underline{b}_4} R_{\underline{b}_1...\underline{b}_4}. \eqno(3) $$

The torsion and the curvatures are restricted by suitable constraints
describing on--shell $D=11$ supergravity.
In particular, (see, for instance, [17]) $$ T_{\underline
{\alpha}\underline{\beta}}{}^{\underline{a}} = 2
\Gamma^{\underline{a}}_{\underline{\alpha}\underline{\beta}} \eqno(4a) $$ $$
R_{\underline{a}\underline{b}\underline{\alpha}\underline{\beta}} =-2i
(\Gamma_{\underline{ab}})_{\underline{\alpha\beta}} \eqno(4b) $$
$$
R_{\underline{a}_1...\underline {a}_5\underline{\alpha}\underline{\beta}} =
-2 (\Gamma_{\underline{a}_1
...\underline{a}_5})_{\underline{\alpha}\underline{\beta}}. \eqno(4c) $$
(In fact, one should not impose these constraints {\sl a
priori}, since they arise as consistency conditions for the five--brane
action to be $\kappa$--invariant.)

For the worldvolume two--form $A_2$ we define the gauge invariant field
strength
$$
H_3 = dA_2 + C_3 \eqno(5)
$$
so that
$$
dH_3 = R_4. \eqno(6)
$$
In eqs. (5) and (6) we imply the pullback of $C_3$ onto the worldvolume.
This pullback will always be understood in what follows.
Our convention for target superforms is $$ \psi_N = {1\over N!}
E^{\underline {A}_1}.. E^{\underline {A}_N} \psi_{\underline
{A}_N...\underline{A}_1} $$ and those in the worldvolume $$ \Phi_n = {1
\over n!} dx^{m_1}...dx^{m_n} \Phi_{m_n...m_1}.  $$ Latin and greek letters
denote respectively vector--like and spinor--like indices, those from the
beginning of the alphabet denote target space indices.
\par
The five--brane
action proposed in [1,2]  contains the fields $Z^M,A_2,$ and a scalar
worldvolume auxiliary field $a(x)$, which insures $d=6$ covariance of the
construction:
$$ I\left[Z,A,a\right]= \int d^6 x \sqrt{-g}
\left({\cal {L}} + {1\over 4} \tilde H^{mn} H_{mn} \right) -
\int \left(C_6 -
{1\over 2} C_3 \wedge H_3\right). \eqno(7) $$
Here $$ g_{mn} (Z) =
E_m{}^{\underline{a}} (Z) E_n{}^{\underline{b}} (Z)
\eta_{\underline{a}\underline{b}} \eqno(8) $$ is the induced worldvolume
metric, which we use to raise and lower (curved) six--dimensional
indices, and $E_m{}^{\underline{A}}= \partial_m
Z^{\underline{M}}E_{\underline{M}}{}^{\underline{A}}$.  ${\cal {L}}$ is the
DBI--like Lagrangian $$ {\cal {L}} = \sqrt{det (\delta_m{}^n + i \tilde
H_m{}^n)}.  \eqno(9) $$ The auxiliary field $a(x)$ enters the action under
derivative. It is convenient to define its vector ``field strength" as
$$ v_m
= {\partial_m a \over \sqrt{-g^{pq}\partial_p a\partial_q a}}\ \ , \eqno(10)
$$ then in (8) $H_{mn}$ and $\tilde H^{mn}$ are defined as follows $$
\eqalignno { H_{mn} &= H_{m n l} v^l,& (11)\cr \tilde H^{mn} &= H^{*mnl}
v_l. &(12)\cr } $$ Note that $v_mv^m=-1$, which allows one to manipulate
with $v_m$ as with a sechsbein component and to essentially simplify many
computations.  The dual of $H$ is defined in the standard way as
$$
H^{*mnl} = {1 \over 3!\sqrt{-g}} \epsilon^{mnlpqr} H_{pqr},
$$ and one has
the identical decomposition $$
H^{mnl}=-{1\over2}{\epsilon^{mnlpqr}\over \sqrt{-g}}v_p\tilde H_{qr}
-3 v^{[m}H^{nl]} \eqno(12a) $$ together with the analogous formula for
$H^*$.  \par The action (7) is invariant under the local symmetries $$
{\delta_1 Z^{\underline{M}} = 0, \qquad \delta_1 a = \varphi(x),
\qquad \delta_1
A_{mn} = {-\varphi \over 2\sqrt{-g^{pq}\partial_p a\partial_q a}}
\left(H_{mn} -{\cal V}_{mn}\right),} \eqno(13) $$
where $$ {\cal V}_{mn} \equiv
-2{\delta {\cal {L}} \over \delta \tilde H^{mn}}, \eqno(14) $$ and $$
{ \delta_2 Z^{\underline{M}} = 0 = \delta_2 a,
\qquad \delta_2 A_{mn} =
\partial_{[m} a~~\Phi_{n]},} \eqno(15) $$ where $\varphi(x)$ and $\Phi_n
(x)$ are infinitesimal transformation parameters, as well as under
the standard gauge symmetry of $A_2(x)$, the $d=6$
diffeomorphisms and $\kappa$--symmetry. The latter is defined by $$
{
\delta_\kappa Z^M E_M{}^{\underline{\alpha}} =
\Delta^{\underline{\alpha}},\qquad  \delta_\kappa Z^M E_M{}^{\underline{a}} =
0,\qquad \delta_\kappa A_2 = - i_\Delta C_3 \qquad \delta_\kappa a=0.}
\eqno(16) $$

Here
$$
\Delta^{\underline{\alpha}} \equiv \left(1 +
\bar\Gamma\right)^{\underline{\alpha}}_{~{\underline{\beta}}}
\kappa^{\underline{\beta}}, \eqno(17) $$
$$
\bar \Gamma = {1 \over {\cal {L}}} \left( \bar \gamma + {i\over 2}
\Gamma^{mnp} v_m \tilde H_{np} - {1\over 16} {\epsilon^{m_1 ... m_6} \over
\sqrt{-g}} \tilde H_{m_1 m_2} \tilde H_{m_3 m_4} \Gamma_{m_5 m_6} \right)
\eqno(18) $$ and
$$
{
\Gamma_m = E_m{}^{\underline{a}} \Gamma_{\underline{a}},\qquad \bar\gamma =
{1 \over 6! \sqrt{-g} }
\epsilon^{m_1 ... m_6}\Gamma_{m_1... m_6}.}
\eqno(19)
$$
The matrix $\bar\Gamma$ satisfies the conditions
$$\bar\Gamma^2 =1, \qquad tr
\bar\Gamma =0. \eqno(20)
$$  Under the $\kappa$--transformations the action
(7) varies as
$$ \delta I = -\int\,\,d^6 x \sqrt{-g} E_m{}^{\underline{\beta}} ({
J}^m)_{\underline{\beta}\underline{\alpha}} \Delta^{\underline{\alpha}}
\eqno(21)
$$ where the matrices ${J}^m$ are $$ {J}^m =
T^{mn}\Gamma_n  + 2 \Gamma^m \bar\gamma + i T^{mnp} \Gamma_{np}, \eqno(22) $$
where $T^{mn}$ is the {\it formal} energy--momentum tensor with respect to
the induced metric (it is not conserved)
$$ T^{mn}={-4\over \sqrt{-g}}{\delta
I\over \delta g_{mn}}= - 2 g^{mn} \left({\cal {L}} - {1\over 2} tr ({\cal
V}\tilde H) \right) +v^m v^n tr ({\cal V} \tilde H)
$$
$$
-2 ({\cal V} \tilde
H)^{mn} -{1\over 2} {v^{(m} \epsilon^{n) p_1..p_5} \over \sqrt{-g}} v_{p_1}
\tilde H_{p_2 p_3} \tilde H_{p_4 p_5} \eqno(23) $$ and
$$ T^{mnp} = 3 v^{[m}
\tilde H^{np]} + {\epsilon^{mnpqrl} \over {2\sqrt{-g}} } v_q {\cal
V}_{rl}.  \eqno(24) $$ Due to the matrix identity $$ {J}^m (1+
\bar\Gamma) =0 \eqno(25) $$ the action is indeed $\kappa$--invariant.
\par The equations of motion of $\vartheta^{\underline{\mu}},
X^{\underline{m}}$, $A_2$ and $a(x)$ are, respectively, $$ \eqalign {
E_m{}^{\underline{\beta}} ({
J}^m)_{\underline{\beta}\underline{\alpha}} &=0,\cr {1\over
2}D_m\left(T^{mn}E_n{}^{\underline{a}}\right)={\epsilon^{m_1\cdots
m_6}\over\sqrt{-g}}&\left({1\over 6!} R^{\underline{a}}{}_{m_6\cdots m_1}
-{1\over (3!)^2} R^{\underline{a}}{}_{m_6 m_5 m_4}H_{m_3 m_2 m_1}\right),
\cr \partial_{[m}\left(v_n\left( {\cal V}_{kl]}-H_{kl]}\right)\right) &=0,
\cr \epsilon^{pqmnkl} \left({\cal V}_{pq}-H_{pq}\right)
\partial_{m}\left(v_n\left( {\cal V}_{kl}-H_{kl}\right)\right)&=0.
}\eqno(25a) $$

Note that the  equation
of motion of $a(x)$ is not independent but is a consequence of the $A_2$
equation.  As has been shown in ref. [1], by appropriately fixing
the gauge transformations in (15) the equation of motion of $A_2$ reduces
to the generalized self--duality condition
$$ H_{mn}={\cal V}_{mn}.\eqno(25b) $$
Using Eq. (25b) and the fact that $\int d^6x\sqrt{-g}\left({\cal {L}}-
{1\over 4} tr\left(H\tilde H\right)\right)$
is $d=6$ diffeomorphism invariant one
can show that $T_{mn}$ satisfies the equation $$ D_mT^{mn}=-{2\over (3!)^2}
{\epsilon^{m_1\cdots m_6}\over\sqrt{-g}} R^n{}_{m_6 m_5 m_4}H_{m_3 m_2 m_1}.
$$
This allows one to rewrite the equation of motion of
$X^{\underline{m}}$ as
$$ \eqalign { {1\over 2}T^{mn}&D_mE_n{}^{\underline{a}}=\cr
&{\epsilon^{m_1\cdots m_6}\over\sqrt{-g}}\left({1\over 6!}
R^{\underline{a}}{}_{m_6\cdots m_1} -{1\over (3!)^2} \left(
R^{\underline{a}}{}_{m_6 m_5 m_4}H_{m_3 m_2 m_1} -E_{n{\underline{a}}}
E^n{}_{\underline{b}} R^{\underline{b}}{}_{m_6 m_5 m_4}H_{m_3 m_2 m_1}
  \right)\right).\cr } \eqno(25c) $$
We conclude the presentation of
the action approach by noting that when the local transformations (13) and
(15) are gauge fixed by the conditions
$$
a(x)= x^5\, \longrightarrow
\,\partial_m a (x) = \delta^5_m; \qquad A_{5m}=0,\eqno(26)
$$
one recovers the formulation of
[3]. In this gauge the invariance under worldvolume diffeomorphisms is no
longer manifest, but still present in a modified form.
\par
Let us now
present the five--brane equations of motion [8,9] which follow from the
doubly supersymmetric geometrical approach [13]. In this formulation the
worldvolume is a supersurface $\Sigma$ locally parametrized by the bosonic
coordinates $x^m$, ($m = 0,...,5$) and the fermionic coordinates
$\vartheta^\mu$, ($\mu = 1,...,16$).  The $Z^{\underline{M}}$ are now
superfields in $\Sigma$, and $A_2$ as well as the pullbacks of target space
superforms become superforms in $\Sigma$.  The essential ingredient of this
approach is the requirement that the embedding of the superworldvolume of the
five--brane into a target superspace respects the condition that the pullback
of $E^{\underline{a}}$ does not have components along the odd directions of
the worldvolume supersurface:  $$ E_\alpha{}^{\underline{a}} =0. \eqno(27) $$

This condition  appeared first in the twistor--like formulation of
superparticles [10] and heterotic strings [11] and it is a characteristic
property of all superbranes in the doubly supersymmetric geometrical
approach [12,13,14].
In many cases, as that of $N=2$, $D=10$ superstrings, $D=11$
supermembranes [12,13] and D--branes [8,14], this condition is so strong that
it forces the model to be on the mass shell. In particular, this is the case
of the M theory five--brane [8,9]. The details of this case
have been worked out in [8,9] and will not be given here. For our
purposes it is sufficient to present the equations of motion for the
worldvolume component fields (which are again $Z^M$ and $A_2$) adapting the
conventions used in those references to ours.
The correct identification of
the worldvolume component fields, which turns out to be a nontrivial
thing, has already been made in [16]. The curvature $H$ and the target
space fields are again defined as in (1)--(6) and all indices are
raised and lowered by the induced metric (8).
\par
The basic auxiliary
field appearing in this approach is an antisymmetric self--dual tensor which
we convert to a tensor with curved indices
$$ h^{mnl} = h^{*mnl} = {1\over
{3!\sqrt{-g}}} \epsilon^{mnlpqr} h_{pqr}.
$$
The consistency of
Bianchi identities implies that this tensor is tied to $H$ through the
relation
$$ 4h_{pqr} = m^l{}_{[p} H_{qr]l}, \eqno(28)
$$
where
$$ \eqalign {
m_{lm} &= g_{lm} + 2 h_l{}^{pq} h_{mpq}\cr &\equiv g_{lm}+2 k_{lm}.\cr } $$
Upon elimination of the auxiliary field $h$, Eq. (28)
turns out
to be the field equation for $H$. With spinor indices suppressed the equation
for $\vartheta^{\underline{\mu}}$ can be written
as $$ E^m\left(1-\gamma^{(3)}\right)\Gamma^n\left(1-\bar
\gamma\right)m_{mn}=0, \eqno(29a) $$ where the matrix $\gamma^{(3)}$ is given
by $$ \gamma^{(3)}=-{i\over 3}h_{lmn}\Gamma^{lmn}= -{i\over
3}h_{lmn}\Gamma^{lmn}\left({1+\bar\gamma\over 2}\right).
$$
Eq. (29a) becomes the
equation of motion of $\vartheta^{\underline{\mu}}$ once, by use of
(28), one expresses
$h$ as a function of $H$, see below.
\par
In [9] the equation of motion of
$X^{\underline{m}}$ is given in the approximation in which one neglects
$E_m{}^{\underline{\alpha}}$ (i.e. drops all terms bilinear (and higher) in
fermions from the bosonic equations)
$$ \eqalign { (m^2)^{mn}
&D_mE_n{}^{\underline{a}}=\cr -&\left(1-{2\over 3}{\it tr}k^2\right)
{\epsilon^{m_1\cdots m_6}\over\sqrt{-g}} \left({1\over 6!}
\hat R^{\underline{b}}{}_{m_6\cdots m_1} -{1\over (3!)^2} \hat
R^{\underline{b}}{}_{m_6 m_5 m_4}H_{m_3 m_2 m_1}\right)
\left(\delta_{\underline b}{}^{\underline a}- E^l{}_{\underline b}
E_l{}^{\underline a}\right).\cr } \eqno(29b) $$
We put the hat on the fields $R_7$ and $R_4$ to remember that the pullback
has been made only with respect to their {\it bosonic} target indices
$\underline{b}$, i.e. with $E_m{}^{\underline b}$.  These component field
equations are covariant under $\kappa$--transformations which are completely
analogous to (16) (together with a suitable transformation law for $h$), but
with the difference that now $$ \Delta^{\underline{\alpha}} = \left(1+
\hat\Gamma\right)^{\underline{\alpha}}{}_{\underline{\beta}}
\hat\kappa^{\underline{\beta}}, \eqno(30)
$$
where
$$ \hat\Gamma = \bar\gamma
+ \gamma^{(3)}.\eqno(31) $$
Notice that $\hat\Gamma$ also satisfies
$\hat\Gamma^2 =1$ and ${\it tr} \hat\Gamma=0$.
\par
The connection between
the two approaches is established as follows.
First, one has to
disentangle the implicit equation (28).  For this define
$$ h_{mn} =
h_{mnp} v^p = h^*_{mnp} v^p \eqno(32) $$
 and use the identity (12a) to project
$H$ and $h$ onto their dual and self--dual parts. One gets
$$
\eqalignno { 4 h_{mn} &= \left(1 -{2\over 3} {\it tr} h^2\right) H_{mn} -
{2\over 3} \left(2 {\it tr} (\tilde H h) h_{mn} - 4 (hh \tilde H )_{mn} - 8
(h h H)_{mn} \right) & (33)\cr 4 h_{mn} &= \left(1+ {2\over 3} {\it tr}
h^2\right) \tilde H_{mn} + {2\over 3} \left(2 {\it tr} (Hh) h_{mn} - 4 (hh
H )_{mn} - 8 (hh \tilde H)_{mn} \right).  & (34)\cr } $$ Despite of a
complicated form, these equations can be solved to get $h_{mn}$ and
$H_{mn}$ in terms of $\tilde H_{mn}$:
$$ \eqalignno { h_{mn}
&= {1\over 4} \tilde H_{mn} + {{1\over 2} ({\it tr} \tilde H^2) \tilde H_{mn}
- 2 \tilde H^3_{mn} \over 8 ({\cal {L}}+1) + 2 {\it tr} \tilde H^2}&
(35)\cr H_{mn} &= {1\over {\cal {L}}} \left((1+ {\it tr} \tilde H^2)\tilde
H_{mn} - \tilde H^3_{mn} \right) = - 2 {\delta L \over \delta \tilde
H^{mn}} ={\cal V}_{mn},&(36)\cr } $$ where ${\cal {L}}$ is the DBI
lagrangian (9). Eq. (37) coincides precisely with (25b), i.e. with the
self--duality condition for $A_2$ in the action approach (see also [16]
where this condition was obtained in a $d=6$ covariant form without any
use of the auxiliary fields).
\par
The
simplest way to obtain Eqs. (35,36) is to use manifest diffeomorphism
invariance and  choose the flat metric $g_{mn}=\eta_{mn}$. Since each of
the antisymmetric matrices $h,H$ and $\tilde H$ live in five dimensions,
following [5] one can perform a five--dimensional Lorentz rotation such
that the only nonvanishing components of $h_{mn}$ are $$ h_{12} = - h_{21}
= h_+, \quad h_{34} = - h_{43} = h_-,
\eqno(37)$$ and similarly for $H_{mn}$ and
$\tilde H_{mn}$. Then Eqs.  (33), (34) become $$ 4 h_\pm = H_\pm \left(1
 \mp 4 (h^2_+ - h^2_-)\right) \eqno(38a) $$ $$ 4 h_\pm = \tilde H_\pm
\left(1\pm 4 (h^2_+ - h^2_-) \right).  \eqno(38b) $$ Eliminating $h_{\pm}$
one gets $$ H_\pm = \tilde H_\pm \sqrt{{1 - \tilde H^2_\mp \over 1 -
\tilde H^2_\pm}} = -{\delta {\cal {L}} \over \delta \tilde H_\pm}={\cal
V}_\pm, \eqno(39) $$ where $$ {\cal
{L}} = \sqrt{(1- \tilde H^2_+) (1- \tilde H^2_-)},\eqno(40) $$
which is the DBI--like Lagrangian in this particular basis. In an analogous
way one can derive (35).
\par
On what concerns the $X$-equation we note
that using (12a) for $h$ we can write the symmetric
matrix $k$ in the form $$ k^{mn} = 4(h^2)^{mn}-g^{mn} {\it  tr} h^2 - 2 v^m
v^n {\it tr}h^2 + {v^{(m} \epsilon^{n) p_1..p_5} \over \sqrt{-g}} v_{p_1}
\tilde h_{p_2 p_3} \tilde h_{p_4 p_5} $$ and use it and (25b) to
derive a remarkable relation $$ (m^2)^{mn}=-{1\over 2}
T^{mn}\left(1-{2\over 3} {\it tr}k^2\right).  $$ If one uses this relation
and drops the terms containing
$E_m{}^{\underline{\alpha}}$ at the r.h.s. of (25c) one gets precisely (29b).
\par The comparison of the $\vartheta$--equations is performed as follows.
First we note that the matrix $\gamma^{(3)}$ can be written as
$$
\gamma^{(3)} = iv_l h_{mn} \Gamma^{lmn}  (1+ \bar\gamma), \eqno(41) $$ so
that $$ \eqalign { 1+ \hat\Gamma &= (1 + \gamma^{(3)}) (1+ \bar\gamma),\cr
1-\hat\Gamma &= (1- \bar\gamma)(1 - \gamma^{(3)}). \cr } \eqno(42) $$ By use
of Eqs. (41), (42),
the definition (18) and (35) it is a simple (but lengthy) exercise to
prove the following matrix identities $$ \eqalign { \left({1 + \bar\Gamma
\over 2}\right) \left({1 + \hat\Gamma \over 2}\right) &= {1 + \hat\Gamma
\over 2},\cr \left({1 + \hat\Gamma  \over 2}\right) \left({1 + \bar\Gamma
\over 2}\right) &= {1 + \bar\Gamma\over 2}. \cr } \eqno(43) $$ A convenient
way to prove these matrix identites is to reduce them to identities
between worldvolume tensors and then to verify the former in the
particular $(\pm)$--basis as above.  The relations (43) show
that the $\kappa$--transformation of the two approaches are the same modulo a
redefinition of the gauge parameter $\kappa$.  Moreover, applying the first
relation in (43)
to the identity (25) one gets $$ {J}^m (1 + \hat\Gamma) =0, \eqno(44) $$
so that the field equation of $\vartheta$ (25a) in the action approach is
written as $$ E_m {J}^m (1 - \hat\Gamma)=0, $$ or, using (42), as
$$ E_m
{J}^m (1 - \bar\gamma) =0. \eqno(45) $$
The last matrix identity required to complete the comparison is
$$ {J}^m (1 - \bar\gamma) = - {4 \over{1 -{2\over3}{\it tr}k^2}}
\left(1-\gamma^{(3)}\right)\Gamma_n\left(1-\bar \gamma\right)m^{mn}
\eqno(46)
$$
To prove this identity one has to use, in particular, the relation $$
T^{mn}-2g^{mn}=-{4 m^{mn}\over 1-{2\over 3}{\it tr}k^2}.
$$
Therefore (45) coincides with (29a) apart from a (nonvanishing) overall
scalar factor.

In conclusion we have shown that the equations of motion of the
M--theory 5--brane obtained from the action principle are identical
to the worldvolume component field equations derived from the doubly
supersymmetric geometrical approach. The identification was established
by solving the relation (28) between the auxiliary self--dual field
$h_{lmn}$ and the field strength $H_{mnl}$ and expressing the former in terms
of the latter (see Eq. (35)). To write this expression in a $d=6$ covariant
way one had to use the auxiliary scalar field $a(x)$.  This resulted in
linking the $X$ and $A_2$ equations of these approaches. Then we found the
$\gamma$--matrix identities (43) which allowed us to relate the parameters
of the $\kappa$--transformations of the action and the geometrical
approach and finally to identify the ${\vartheta}$-equations.

A direction of further work might be studying the possibility of getting
the complete set of the superfield equations of the M--theory five--brane
[8] from a generalized action principle proposed in [13] for the
description of
superbranes in the doubly supersymmetric approach. This has already been
done for $D=10$ D-branes in [14].

One might also try to look for a 5--brane action which includes the
field $h_{mnl}$ instead of $a(x)$. Perhaps it would involve a
covariant formulation of self--dual field dynamics (with infinite
number of auxiliary (anti)--self--dual fields) developed in [18].

\vskip 0.3truecm
{\bf Acknowledgements}. Authors are grateful to I. Arefieva, E.
Bergshoeff, T. Ortin and M. Vasiliev for discussion.  Work of K.L.,
P.P.  and M.T.  was supported by the European Commission TMR programme
ERBFMRX--CT96--045 to which K.L., P.P. and M.T. are associated. I.B.,
A.N. and D.S.  acknowledge partial support from grants of
the Ministry of Science and Technology of Ukraine and the INTAS Grants
N 93--127, N 93--493--ext, and N 94--2317.

\vskip 0.3truecm

{\bf REFERENCES}

\vskip 0.3truecm
\item{[1]}
P. Pasti, D. Sorokin and M. Tonin, Covariant action for a D=11 5--brane
with the chiral field, hep--th/9701037 ({\sl Phys. Lett.} {\bf B} in
press).
\item{[2]} I. Bandos, K. Lechner, A. Nurmagambetov, P. Pasti, D. Sorokin and
M.  Tonin, Covariant Action for the Super--Five Brane of M--theory,
preprint DFPD97/TH/05, hep--th/9701149.
\item{[3]} M. Aganagic, J. Park,
C. Popescu, and J.H. Schwarz, World--Volume Action of the M Theory
Five--Brane, hep--th/9701166.
\item{[4]} P. K. Townsend, {\sl Phys.
Lett.} {\bf B373} (1996) 68.

O. Aharony, String theory dualities from M--theory, hep--th/9604103.

E. Bergshoeff, M. de Roo and  T. Ortin, The eleven--dimensional
five--brane, hep--th/9606118.

E. Witten, Five--brane effective action in M theory, hep--th/9610234.

\item{[5]}
 M. Perry and J. H. Schwarz, Interacting chiral gauge fields in six
dimensions and Born--Infeld theory, hep--th/9611065.

J.H. Schwarz, Coupling a Self--Dual Tensor to Gravity in Six Dimension,
hep--th/9701008.
\item{[6]}
P. Pasti, D. Sorokin, and M. Tonin, {\sl Phys. Lett.} {\bf
B352} (1995) 59;

P. Pasti, D. Sorokin and M. Tonin, {\sl Phys. Rev.} {\bf D52} (1995) R4277;

P. Pasti, D. Sorokin and M. Tonin, On Lorentz Invariant Actions for Chiral
P--Forms, hep--th/9611100 ({\sl Phys. Rev.} {\bf D} in press).
\item{[7]} S. Cherkis and J. H. Schwarz, Wrapping the M theory Five--Brane
on K3, hep--th/9703062.
\item{[8]} P.S. Howe and E. Sezgin, Superbranes, hep--th/9607227.

P.S. Howe and E. Sezgin, D=11, p=5, hep--th/9611008.
\item{[9]} P.S. Howe, E. Sezgin and P. West, Covariant field equations of the
M theory five--brane, hep--th/9702008.
\item{[10]} D. Sorokin, V. Tkach
and D. V. Volkov, {\sl Mod. Phys. Lett.} {\bf A4} (1989) 901.

D. Sorokin, V. Tkach, D. V. Volkov and A. Zheltukhin, {\sl Phys. Lett.}
{\bf B216} (1989) 302.

A. Galperin and E. Sokatchev, {\sl Phys. Rev.} {\bf D46} (1992) 714.
\item{[11]}
N. Berkovits, {\sl Phys. Lett.} {\bf 232B} (1989) 184;

M. Tonin, {\sl Phys. Lett.} {\bf B266} (1991) 312;
{\sl Int. J. Mod. Phys.} {\bf 7} (1992) 613;

S. Aoyama, P. Pasti and M. Tonin, Phys. Lett. {\bf B283} (1992) 213.

F. Delduc, A. Galperin, P. Howe and E. Sokatchev, {\sl Phys. Rev.}
{\bf D47} (1992) 587.
\item{[12]}
A. Galperin and E. Sokatchev, {\sl Phys. Rev.} {\bf D48} (1993) 4810.

P. Pasti and M. Tonin, {\sl Nucl. Phys.} {\bf 418} (1994) 337.

E. Bergshoeff and E. Sezgin, {\sl Nucl. Phys.} {\bf B422} (1994) 329.
\item{[13]}
I. Bandos, P. Pasti, D. Sorokin, M. Tonin and D. Volkov,
{\sl Nucl. Phys.} {\bf B446} (1995) 79.

I. Bandos, D. Sorokin and D. Volkov,
{\sl Phys. Lett.} {\bf B352} (1995) 269.
\item{[14]}
I. Bandos, D. Sorokin and M. Tonin, Generalized action principle and
superfield equations of motion for D=10 D--p--branes,
hep--th/9701127.
\item{[15]} M. Aganagic, J.Park, C. Popescu and J.H. Schwarz, Dual D--Brane
Actions,
hep--th/9702133.
\item{[16]} P.S. Howe, E. Sezgin and P. West, The six--dimensional self--dual
tensor, hep--th/9702111.
\item{[17]} A. Candiello and K. Lechner,
{\sl Nucl. Phys.} {\bf B412} (1994) 479.
\item{[18]}
B. McClain, Y. S. Wu, F. Yu, {\sl Nucl. Phys.} {\bf B343} (1990) 689.

C. Wotzasek, {\sl Phys. Rev. Lett.} {\bf 66} (1991) 129.

I. Martin and A. Restuccia, {\sl Phys. Lett.} {\bf B323} (1994) 311.

F. P. Devecchi and M. Henneaux, {\sl Phys. Rev.} {\bf D45} (1996) 1606.

N. Berkovits, Manifest electromagnetic duality in closed superstring
theory, hep-th/9607070.

I. Bengtsson and A. Kleppe, On chiral p--forms, hep-th/9609102.
\bye